\documentclass{aipproc}
\usepackage{epsfig}
\usepackage{axodraw}
\usepackage{epsf}
\usepackage{fancybox}
\usepackage{colordvi}
\usepackage{amsmath,feynmp,emp,color}
\newcommand{\bi}{\begin{itemize}}
\newcommand{\ei}{\end{itemize}}
\begin{document}

\title{Gauge Invariant Classes of Feynman Diagrams and Applications for
Calculations}

\author{E.E.~Boos}

\affiliation{Skobeltsyn Institute of Nuclear Physics, 
            Moscow State University, Moscow 119899, Russia}

\begin{abstract}
In theories like SM or MSSM with a complex gauge group structure  the
complete set of Feynman diagrams contributed to a particular physics
process can be splited to exact gauge invariant subsets.
Arguments and examples given in the review demonstrate that 
in many cases computations and analysis of the gauge
invariant subsets are important.
\end{abstract}

\maketitle

The increase of collider energies requires computations of
processes with more particles in the final state and with better
precision (NLO, NNLO etc). At LEP1 the basic processes were
2 fermions ($\gamma$) production; LEP2 deals basically with
4 fermion ($\gamma$) processes; Tevatron, LHC and LC in many cases
need in an analysis of the processes with 5,6,8 and so on fermions in the
final state, for example,
top pair production with decays - 6 fermions;
single top production in W-gluon fusion mode - 5 fermions;
strongly interacting Higgs sector in hadronic collisions-
$p p \rightarrow q \bar{q} W^+W^-$ - 6 fermions;
study of Yukawa coupling in $p p (e^+e^-) \rightarrow t \bar{t} H$
- 8 fermions etc.
Typically for the processes with multi particle final states
 a number of contributing diagrams is large.
For hadronic collisions not only a number of diagrams but
also a number of partonic subprocesses is very large.

 One of the problem in process computations is a gauge
cancellation among
many of diagrams. Any method of calculation should preserve gauge
invariance which is rather complicated in theories like SM with
not a simple gauge group.
 The well known statement from the quantum theory of gauge fields is that the
whole set of Feynman diagrams contributing to any physics process is exactly 
gauge invariant.
However, in practical calculations it remains amazing how gauge
cancellations take place. But in general the
complete set of Feynman diagrams contributed to a particular physics
process can be splited to exact gauge invariant subsets, and
the gauge cancellations occur in each of the subset.
It will be demonstrated that in many cases the idea to split 
the complete set of diagrams on exact gauge invariant subsets 
\cite{boos-ohl} could be useful in practice:
\bi
\item{ any physics approximation should be based on gauge invariant classes of
diagrams}
\item{ better precision of computations in many cases }
\item{ better understanding of physics parameters like running couplings
or scales (QCD scale, ISR scale etc).  Often different part 
of the same process need
in different values because different kinematical regions might be important}
\item{ any MC generator needs in 
well behaved and compact matrix elements and the gauge invariance gives that}
\ei
 For simple cases it is very easy to find gauge invariant subclasses of diagrams.
Take, for instance, the Bhabha-scattering. In that case the
two s-channel and two t-channel SM diagrams are separately gauge invariant
 as immediately
follows if one substitutes the final $e^+e^-$ pair by the $\mu^+\mu^-$ pair.
In fact, it gives the simplest example of so called "Flavor Flip" introduced
in \cite{boos-ohl}. There are two kinds of flips: Gauge and Flavor,
which correspond to permutations the $2\rightarrow2$ suddiagrams with
on- and/or off-shell legs 
(see the exact definitions of flips in \cite{boos-ohl}).
For a concrete physical process one can get one Feynman diagram from another
by a sequence of gauge and flavor flips. In correspondence to all the diagrams
for the process one may put some graph called Forest.  
The  Forest is a graph with each vertex representing
a diagram and the edges given by the flips (gauge and/or flavor) of four-point
sub-diagrams. The Gauge Forest is such a Forest
or part of the Forest in which the points connected by the only gauge flips.
The connected components of the Gauge Forest are called Groves.
The general theorem has been proved in \cite{boos-ohl} by the 
mathematical induction method:

The Forest F(E) for an external state E consisting of gauge and
matter fields is connected if the fields in E carry no conserved 
quantum numbers other than the gauge charges. 
The Groves are the minimal gauge
invariant classes of Feynman diagrams.

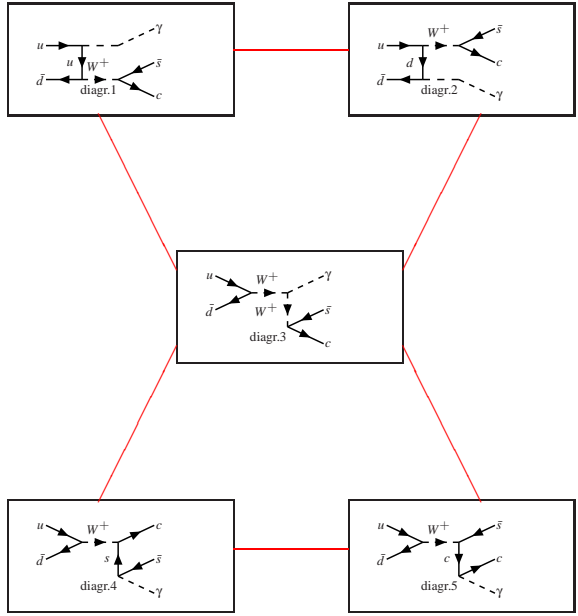
\begin{figure}[htbp]
{
\unitlength=0.8pt
\SetScale{0.8}
\SetWidth{0.7}      
\tiny    
{} \qquad\allowbreak
\begin{picture}(240,290)
  \put(0,235){\fbox{\parbox[b]{79pt}{
    \begin{picture}(79,45)(0,20)
    \Text(13.0,49.0)[r]{$u$}
    \ArrowLine(14.0,49.0)(31.0,49.0) 
    \DashLine(31.0,49.0)(48.0,49.0){3.0} 
    \Text(66.0,57.0)[l]{$\gamma$}
    \DashLine(48.0,49.0)(65.0,57.0){3.0} 
    \Text(27.0,41.0)[r]{$u$}
    \ArrowLine(31.0,49.0)(31.0,33.0) 
    \Text(13.0,33.0)[r]{$\bar{d}$}
    \ArrowLine(31.0,33.0)(14.0,33.0) 
    \Text(39.0,37.0)[b]{$W^+$}
    \DashArrowLine(31.0,33.0)(48.0,33.0){3.0} 
    \Text(66.0,41.0)[l]{$\bar{s}$}
    \ArrowLine(65.0,41.0)(48.0,33.0) 
    \Text(66.0,25.0)[l]{$c$}
    \ArrowLine(48.0,33.0)(65.0,25.0) 
    \Text(39,25)[b] {diagr.1}
    \end{picture}}}}
  \put(161,235){\fbox{\parbox[b]{79pt}{
    \begin{picture}(79,45)(0,20)
    \Text(13.0,49.0)[r]{$u$}
    \ArrowLine(14.0,49.0)(31.0,49.0) 
    \Text(39.0,53.0)[b]{$W^+$}
    \DashArrowLine(31.0,49.0)(48.0,49.0){3.0} 
    \Text(66.0,57.0)[l]{$\bar{s}$}
    \ArrowLine(65.0,57.0)(48.0,49.0) 
    \Text(66.0,41.0)[l]{$c$}
    \ArrowLine(48.0,49.0)(65.0,41.0) 
    \Text(27.0,41.0)[r]{$d$}
    \ArrowLine(31.0,49.0)(31.0,33.0) 
    \Text(13.0,33.0)[r]{$\bar{d}$}
    \ArrowLine(31.0,33.0)(14.0,33.0) 
    \DashLine(31.0,33.0)(48.0,33.0){3.0} 
    \Text(66.0,25.0)[l]{$\gamma$}
    \DashLine(48.0,33.0)(65.0,25.0){3.0} 
    \Text(39,25)[b] {diagr.2}
    \end{picture}}}}
  \put(80,118){\fbox{\parbox[b]{79pt}{
    \begin{picture}(79,45)(0,20)
    \Text(13.0,57.0)[r]{$u$}
    \ArrowLine(14.0,57.0)(31.0,49.0) 
    \Text(13.0,41.0)[r]{$\bar{d}$}
    \ArrowLine(31.0,49.0)(14.0,41.0) 
    \Text(39.0,53.0)[b]{$W^+$}
    \DashArrowLine(31.0,49.0)(48.0,49.0){3.0} 
    \Text(66.0,57.0)[l]{$\gamma$}
    \DashLine(48.0,49.0)(65.0,57.0){3.0} 
    \Text(44.0,41.0)[r]{$W^+$}
    \DashArrowLine(48.0,49.0)(48.0,33.0){3.0} 
    \Text(66.0,41.0)[l]{$\bar{s}$}
    \ArrowLine(65.0,41.0)(48.0,33.0) 
    \Text(66.0,25.0)[l]{$c$}
    \ArrowLine(48.0,33.0)(65.0,25.0) 
    \Text(39,25)[b] {diagr.3}
    \end{picture}}}}
  \put(0,0){\fbox{\parbox[b]{79pt}{
    \begin{picture}(79,45)(0,20)
    \Text(13.0,57.0)[r]{$u$}
    \ArrowLine(14.0,57.0)(31.0,49.0) 
    \Text(13.0,41.0)[r]{$\bar{d}$}
    \ArrowLine(31.0,49.0)(14.0,41.0) 
    \Text(39.0,53.0)[b]{$W^+$}
    \DashArrowLine(31.0,49.0)(48.0,49.0){3.0} 
    \Text(66.0,57.0)[l]{$c$}
    \ArrowLine(48.0,49.0)(65.0,57.0) 
    \Text(44.0,41.0)[r]{$s$}
    \ArrowLine(48.0,33.0)(48.0,49.0) 
    \Text(66.0,41.0)[l]{$\bar{s}$}
    \ArrowLine(65.0,41.0)(48.0,33.0) 
    \Text(66.0,25.0)[l]{$\gamma$}
    \DashLine(48.0,33.0)(65.0,25.0){3.0} 
    \Text(39,25)[b] {diagr.4}
    \end{picture}}}}
  \put(161,0){\fbox{\parbox[b]{79pt}{
    \begin{picture}(79,45)(0,20)
    \Text(13.0,57.0)[r]{$u$}
    \ArrowLine(14.0,57.0)(31.0,49.0) 
    \Text(13.0,41.0)[r]{$\bar{d}$}
    \ArrowLine(31.0,49.0)(14.0,41.0) 
    \Text(39.0,53.0)[b]{$W^+$}
    \DashArrowLine(31.0,49.0)(48.0,49.0){3.0} 
    \Text(66.0,57.0)[l]{$\bar{s}$}
    \ArrowLine(65.0,57.0)(48.0,49.0) 
    \Text(44.0,41.0)[r]{$c$}
    \ArrowLine(48.0,49.0)(48.0,33.0) 
    \Text(66.0,41.0)[l]{$c$}
    \ArrowLine(48.0,33.0)(65.0,41.0) 
    \Text(66.0,25.0)[l]{$\gamma$}
    \DashLine(48.0,33.0)(65.0,25.0){3.0} 
    \Text(39,25)[b] {diagr.5}
    \end{picture}}}}
  \put(107,262){{\color{red}\line(1,0){54}}}
  \put(43,232){{\color{red}\line(1,-2){37}}}
  \put(224,232){{\color{red}\line(-1,-2){37}}}
  \put(107,26){{\color{red}\line(1,0){54}}}
  \put(43,48){{\color{red}\line(1,2){37}}}
  \put(224,48){{\color{red}\line(-1,2){37}}}
\end{picture}}
  \caption{\label{fg:2}%
    The Forest for the process $ u \bar{d} \rightarrow c \bar{s} \gamma $
     is the Grove.}
\end{figure}

A very simple forest as an example is shown in Fig.~\ref{fg:2} 
 for the process $ u \bar{d} \rightarrow c \bar{s} \gamma $.
All diagrams are connected by the gauge flips passing through
the middle diagram 3. Therefore all of them form 
one Grove - one gauge invariant set of diagrams. 

More complex examples of the Forests and 
Groves are shown in \cite{boos-ohl}. They have been obtained by means 
of the program \textit{bocages} \cite{ohl} in which the algorithm
of tree diagrams generation based on the flips has been realised.
 Table  showns how the number of  diagrams for various processes with 6
external fermions  splits into minimal gauge invariant subsets (Groves).
  \begin{tabular}{l|r|l}
    $E$ & $\sum$ & classes \\\hline
    $u \bar u u \bar u u \bar u$
      & $144$ & $18\cdot8$ \\
    $u \bar u u \bar u u \bar u\gamma$
      & $1008$ & $18\cdot24+36\cdot16$ \\
    $u \bar u u \bar u d \bar d$
      & $92$ & $4\cdot{\textRubineRed 11}\textBlack+6\cdot8$ \\
    $u \bar u u \bar u d \bar d\gamma$
      & $716$ & $4\cdot95+6\cdot24+12\cdot16$ \\   
    $\ell^+ \ell^- u \bar u d \bar d$
      & $35$ & $1\cdot\textRubineRed 11 \textBlack +3\cdot8$ \\
    $\ell^+ \ell^- u \bar u d \bar d\gamma$
      & $262$ & $1\cdot94+3\cdot24+6\cdot16$ \\  
    $\ell^- \nu d \bar u d \bar d$
      & $20$ & $2\cdot\textRubineRed 10 \textBlack$\\
    $\ell^- \nu d \bar u d \bar d\gamma$
      & $152$ & $2\cdot76$ \\
    $\ell^+ \ell^- \ell^- \nu d \bar u$
      & $20$ & $2\cdot\textRubineRed 10 \textBlack$\\
    $\ell^+ \ell^- \ell^- \nu d \bar u\gamma$
      & $150$ & $2\cdot75$ \\
    $\ell^- \nu \ell^+ \bar\nu d \bar d$
      & $19$ & $1\cdot\textRubineRed 9 \textBlack +2\cdot4+1\cdot2$ \\
    $\ell^- \nu \ell^+ \bar\nu d \bar d\gamma$
     & $107$ & $1\cdot59+2\cdot12+2\cdot8+2\cdot4$ \\
    $\ell^- \bar\nu \ell^+ \nu \ell^+ \ell^-$
      & $56$ & $4\cdot\textRubineRed 9 \textBlack +4\cdot4+2\cdot2$ \\
    $\ell^- \bar\nu \ell^+ \nu \ell^+ \ell^-\gamma$
      & $328$ & $4\cdot58+4\cdot12+4\cdot8+4\cdot4$ \\
     $\ell^+ \nu \ell^- \bar\nu \nu \bar\nu$
      & $36$ & $4\cdot6+6\cdot2$ \\
    $\ell^+ \nu \ell^- \bar\nu \nu \bar\nu\gamma$
      & $132$ & $4\cdot26+2\cdot6+4\cdot4$ \\
    $\nu \bar\nu \nu \bar\nu \nu \bar\nu$  
      & $36$ & $18\cdot2$
  \end{tabular}
The familiar LEP2 gauge invariant classes (\textRubineRed{CC09},
\textRubineRed{CC10}, \textRubineRed{CC11}, \textBlack etc.)
 appear here automatically. However now we know that these
 classes are not only gauge invariant but they are minimal classes.
 
Let us take as an example the CC20 process (so called "single W")
$e^+ e^- \rightarrow e^- \nu d \bar{u}$ which splits to t- and s-channel
CC10 gauge invariant subclasses. By means of the CompHEP
\cite{comphep} one can compute the contributions of the classes and
their interference as shown in the Table below \cite{boos-dubinin}
(quark phase space cuts: $E_q \ge$ 3 GeV, $M_{ud} \ge$ 5 GeV  and the
lepton phase space cut:  $\cos \theta_e \ge$ 0.997).
\begin{tabular}{|c|c|c|c|}
\hline
$\sqrt{s}$ & $\sigma(CC10-t)$ & $\sigma(CC10-s)$ & $\sigma(t-s interf.)$ 
\\ \hline
\multicolumn{4}{|c|}{quark phase space cuts, no ISR} \\ \hline
190      &\bf 147(0)         &\bf 680(1)         &\bf 5(0)     \\
350      &\bf 635(1)         &\bf 420(1)         &\bf 21(0)    \\
500      &\bf 1127(2)        &\bf 270(0)         &\bf 19(0)    \\
800      &\bf 1981(4)        &\bf 143(0)         &\bf 16(0)    \\
\hline
\multicolumn{4}{|c|}{lepton and quark phase space cuts, no ISR} \\ \hline
190      &\bf 116(0)         &\bf 2(0)          &\bf 0.0(0)    \\
350      &\bf 513(1)         &\bf 7(0)          &\bf 0.2(0)     \\
500      &\bf 928(2)         &\bf 10(0)         &\bf 0.3(0)     \\
800      &\bf 1671(4)        &\bf 15(0)         &\bf 0.4(0)    \\
\hline
\end{tabular}
The CC10 t-channel part contains the single W boson production,
it grows up with the collision energy and it starts to dominate the
CC10 s-channel part (W boson pair production) at about 320 GeV. 
One should stress few points here
\bi
\item{ a good precision of computations is obtained 
 only if one splits the complete CC20 set of diagrams to CC10 subsets
 because in that case one can use different kinematical variables of
 integration for different subsets with different mapping of singularities
 (the interference contribution is small)}
\item{ the "overall" scheme of the W-boson width treatment could be used
only for separate classes, otherwise there will be an artificial 
suppression of the CC10 t-channel part by the factor related to the second  
W pole}
\item{ for the CC10 s-channel part obviously the scale of the order of 
energy should be used for the electro-magnetic $\alpha$ and ISR
while the CC10 t-channel part has a very small characteristic virtuality
of the soft virtual photon, and therefor a typical scale for the 
corresponding $\alpha$ and ISR should be taken much smaller of the order
electron momentum transfer \cite{boos-dubinin, kurihara}}
\ei

The same general statements are also true for the process of
so called "single Z" production 
 $e^+ e^-   \rightarrow e^+ e^- \nu \bar{\nu}$.
 In this case there are  56 Feynman diagrams 
which split to 10 gauge invariant classes or 10 Groves
$56 D =  4 * 9D + 4 * 4D + 2 * 2D$ \cite{boos-dubinin2}.
Two classes of 4 diagrams each contribute to the single Z 
as given in the second column of the Table below.
In the Table the contributions of the gauge invariant subsets 
in fb are given at the energy
$\sqrt{s}=$200 GeV. First row - with angular cuts, second row - no
angular cuts for $e^-$, $e^+$. The following 
angular and lepton energy cuts are used: 
 $\cos \theta_{e^-} \ge$ 0.997 and $\cos \theta_{e^+} \le$
0.997 and $E_l \ge$ 15 GeV.
\begin{tabular}{|c|c|c|c|c|} \hline
 &18$W$ & 8$Z$ & 9$W^+W^-$ & 4$ZZ$
\\ \hline
$\theta_e$,$E_l$
 & 36.1& 16.4& 0.91& 0.02
\\ \hline  
only $E_l$
 &106.6 & 153.6 &240.5 &44.9
\\ \hline
\end{tabular}

If the energy and angular cuts are applied still there are significant
contributions from both single W (first column) and Z (second column) 
productions. 
One should be
careful in interpretation of experimental measurements in this channel.
(Contributions of other gauge invariant subsets are very small and we
do not show them here). 


One more example of applications 
of the gauge classes  is related  to 
the method of simplification of flavour combinatorics
        in the evaluation of hadronic processes \cite{boos-ilyin}. 
Here a serious computational problem  
 is the large number of partonic subprocesses
due to a presence of many quark partons with different flavors
 in the colliding hadrons
and contributions of many additional diagrams 
for each subprocess because of the CKM quark mixing.
However in the approximation when 
CKM matrix is reduced to the CK matrix without a mixing with 
the 3d quark generation 
\vspace*{-0.3cm}
$$ V_{CKM}  \;\Longrightarrow\; \left( \begin{array}{cc}
                                            V & 0 \\
                                            0  & 1
                                       \end{array} \right)\;,   
   \qquad
   V \;=\;  \left( \begin{array}{cc}
                         \cos\vartheta_c  & \sin\vartheta_c \\
                         -\sin\vartheta_c & \cos\vartheta_c
                                       \end{array} \right)$$
where $\vartheta_c$ is the Cabbibo angle
and neglecting masses of the quarks from the first two generations  
$M_u = M_d = M_s = M_c = 0$ the problem can be simplified drastically.
In this case diagrams contributed to the process can be splited to
the gauge invariant classes with different topologies of the incoming
and outgoing quark lines. Then one can make a rotation of down quarks
in all vertices of Feynman diagrams 
thus, transporting the mixing matrix elements 
from the diagrams to the parton distribution
functions . As a result a number
of rules for a convolution with quark distribution functions appear
depending on the topology of the gauge invariant class
(see details in \cite{boos-ilyin}, the method has been realized in 
the CompHEP version  for hadron collisions V41.10). 

In this talk we have discussed the method which allows 
to split the complete set
of Feynman diagrams contributed to a physics process on gauge
invariant subclasses. Well known gauge invariant classes of diagrams like
CC10, CC11, CC09 etc naturally appear in such an approach.
It was demonstrated the above classes are the minimal invariant classes
("Groves").
For a concrete physical process one creates the graph -
"Forest" in which vertices represent diagrams and edges
show the connection between diagrams by possible flips,
flavor and gauge. The vertices of the graph (diagrams) connected by the only
gauge flips form connected subgraphs, "Groves", and the corresponding
diagrams form minimal gauge invariant classes.
The flavor flips connect diagrams from different gauge
invariant classes.

Separation on gauge invariant classes in some cases
allows to understand better properties of processes,
to get better precision of calculations, to make
in tree level computations a natural choice
of characteristic scales for ISR, structure functions, running
couplings etc, to get reasonable approximations, 
it leads to a simplification of flavor combinatorics etc.

In some cases for processes with multi-particle final states
the number of gauge invariant classes is much smaller than
the number of physical reactions \cite{boos-ohl2}. So one can compute,
in principle,
amplitudes for gauge invariant subclasses of diagrams and
then compute processes by taken different combinations of that
amplitudes for classes.

The analysis was done for the tree level Feynman
diagrams. A consideration at loop level is in progress \cite{boos-ohl3}. 


The author thanks the Organising Committee of the ACAT2000 Conference 
for kind hospitality. 
I would like to thank my collaborators and coauthors
 T.~Ohl, M.~Dubinin, and V.~Ilyin. 
The work was partly supported by the RFBR-DFG 99-02-04011,
RFBR 00-01-00704, CERN-INTAS 99-377, and Universities of Russia
990588 grants.
\vspace*{-0.3cm}
\nocite{*}
\bibliographystyle{aipproc}

\end{document}